\documentclass[aps,reprint,groupaddress]{revtex4-1}
\usepackage{amssymb}
\usepackage{amsmath}
\usepackage{dcolumn}
\usepackage{bm}
\usepackage[dvips]{epsfig}
\usepackage{graphicx}
\usepackage{mathrsfs}
\usepackage[colorlinks,linkcolor=blue,anchorcolor=blue,citecolor=blue,urlcolor=black]%
{hyperref}

\begin{document}

\title{Enhanced electromechanical coupling of a nanomechanical resonator  to coupled superconducting cavities }
\author{Peng-Bo Li}
\email{lipengbo@mail.xjtu.edu.cn}
\author{Hong-Rong Li}

\author{Fu-Li Li}
\affiliation {Institute of Quantum Optics and Quantum Information,
Department of Applied Physics, Xi'an Jiaotong University, Xi'an
710049, China}

\begin{abstract}
\textbf{We investigate the electromechanical coupling between a nanomechanical resonator and two parametrically coupled superconducting coplanar waveguide cavities that are driven by a two-mode squeezed microwave source. We show that, with the selective coupling of the resonator to the cavity Bogoliubov modes,    the radiation-pressure type coupling   can be greatly  enhanced by several orders of magnitude, enabling  the single photon strong coupling to be reached. This  allows the investigation of a number of interesting phenomena such as photon blockade effects and the generation of nonclassical quantum states  with electromechanical systems.}
\end{abstract}

\maketitle

Cavity optomechanics \cite{sci-321-1172,phys-2,RMP-86-1391} and electromechanics \cite{jpcs-264,Nat-471-204,crp-13-470} are  pretty promising  for fundamental studies of  large-scale quantum phenomena
as well as  appealing applications in quantum science and technology.
Recent experimental progresses have demonstrated ground state cooling of the mechanical resonators \cite{Nat-475-359,Nat-478-89}, coherent coupling between cavity and mechanical modes \cite{Nat-482-63,Nat-495-210,Natphs-9-179}, optomechanically induced transparency \cite{Sci-330-1520,Nat-472-69}, and the generation of squeezed light \cite{Nat-488-476,Nat-500-185}. In despite of these remarkable advances, however, there is a serious hindrance in this exciting field, i.e.,  the radiation-pressure coupling is too weak to ensure dynamics of the system in the single photon strong coupling regime. Current experiments have mainly relied on strong optical driving, which enhances the coupling at the expense of making the effective interaction linear. As a result, this linear optomechanical interaction does not possess the ability of generating nonclassical states or give rise to true photon-photon interactions for implementing single-photon quantum processes.

To give a better understanding and fully exploit the regime of strong radiation pressure coupling, it is highly desirable to find an efficient method  for
realizing the strong nonlinear interaction between the vibrations and the electromagnetic field in a realistic setup. Such a regime is particularly important to
test the fundamental theory of quantum physics and to explore potential applications of optomechanical or electromechanical devices to future quantum technology \cite{pra-51-2537,pra-56-4175,prl-107-063601,prl-107-063602,prl-109-013603,prl-109-063601,SR-3-2943,SR-4-6302,prx-1-021011,prl-111-083601,
prl-111-053602,prl-111-053603,prl-112-150602,pra-85-053832,pra-87-053849}.
Some recent theoretical proposals for entering the strong
coupling regime include the use of collective effects in arrays of mechanical oscillators \cite{prl-109-223601} and Kerr nonlinearity via the Josephson effect \cite{prl-112-203603,njp-16-072001}, as well as  the usage of  an
inductive coupling to a flux-dependent quantum circuit \cite{prl-114-143602}.

Recently,  a proposal using
the squeezed optical cavity field to enhance the nonlinear
coupling in an optomechanical system has been introduced \cite{prl-114-093602}. In that proposal, the single-mode cavity field is squeezed by an optical $\chi^{(2)}$ nonlinear medium, while the broadband-squeezed vacuum is
introduced to suppress the noise of the squeezed cavity mode. Though this protocol seems promising,
it can not be straightforwardly applied to electromechanical systems. First, that proposal needs an optical
nonlinear crystal possessing a large $\chi^{(2)}$ nonlinearity, which is quite demanding and technically challenging in the field of electromechanics with microwave frequencies. Secondly, in order to have implications in quantum information science, it is desirable to consider multi-cavities rather than a single cavity for the purpose of distributed quantum computation and quantum network \cite{pra-79-042339}. However, the generalization of this model to the case of coupled optical cavities is still difficult. Fortunately, the above issues can be overcome by considering
two coupled superconducting coplanar waveguide (CPW) cavities with the parametrical coupling form  \cite{prb-87-014508,prl-113-093602}.

In this work, we investigate an electromechanical system consisting of a nanomechanical resonator capacitively interacting with two parametrically coupled CPW cavities. We show that, when the cavities  are driven by a  spectrally broadband two-mode squeezed vacuum,   the radiation-pressure type coupling  of the resonator to the cavity Bogoliubov modes can be greatly enhanced by several orders of magnitude.  By suitably tuning
the system parameters such as the squeezing parameter of the driving source, the flux driving frequency and the parametric coupling strength, the single photon
electromechanical coupling strength can be tailored such that it can exceed the cavity decay rate. This single photon strong coupling of electromechanical interactions allows the studies of single-photon quantum processes such as photon blockade and the production of nonclassical photon states harnessing the optomechanical nonlinearity. With currently available technology in cavity electromechanics, this proposal can be realistically implemented in experiments and
provides an appealing platform for  implementing quantum technologies.

\begin{figure}[b]
\centerline{\includegraphics[bb=53 266 916 558,totalheight=1.1in,clip]{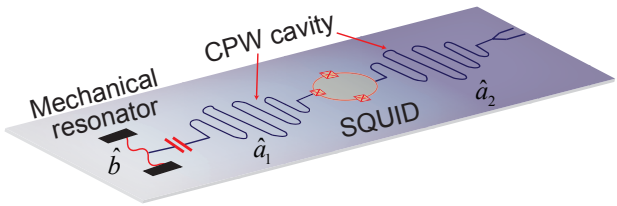}}
\caption{(Color online) Schematic of the proposed  electromechanical setup. A nanomechanical resonator capacitively couples to  a superconducting CPW cavity, which is also coupled to another cavity with the form of parametric coupling. These cavities  are driven by a  spectrally broadband two-mode squeezed vacuum from the output of a non-degenerate Josephson
parametric amplifier. }
\end{figure}
\noindent\textbf{Results}\\
\noindent\textbf{The setup.} As shown in Fig. 1, we consider an electromechanical   system consisting of
a nanomechanical resonator capacitively coupled to  a superconducting CPW cavity, which is also coupled to another auxiliary waveguide  cavity  by means of a
superconducting quantum interference device (SQUID) \cite{prb-87-014508,prl-113-093602}.  The SQUID driven by external fluxes allows a fast modulation of the electrical boundary
condition of the cavities and their interaction \cite{prb-87-014508,prl-113-093602}. In a recent experiment, it has demonstrated a $3$ D
microwave superconducting cavity parametrically coupled to a transmission line cavity by a Josephson ring
modulator \cite{prl-114-090503}.
The mechanical resonator couples to the cavity field via radiation-pressure coupling, while the two cavities interact with each other with the form of parametric coupling. In the frame rotating at half the flux driving
frequency $\omega_\text{d}$, the system Hamiltonian reads
\begin{eqnarray}
  \hat{\mathcal{H}} &=& \hbar \Delta_1 \hat{a}^\dag_1\hat{a}_1+\hbar \Delta_2 \hat{a}^\dag_2\hat{a}_2+\hbar \xi  (\hat{a}^\dag_1\hat{a}_2^\dag+\hat{a}_1\hat{a}_2)\nonumber\\
  &&+\hbar\omega_\text{m}\hat{b}^\dag\hat{b}+\hbar g\hat{a}^\dag_1\hat{a}_1(\hat{b}^\dag+\hat{b}),
\end{eqnarray}
where $\hat{a}_j$ is the annihilation operator for the $j$th cavity mode with frequency $\omega_j$, $\Delta_j=\omega_j-\omega_\text{d}/2$,
$\xi$ the parametric coupling strength between the cavities, $\omega_\text{m}$ the mechanical vibration frequency with annihilation operator $\hat{b}$,
and $g$ the radiation-pressure coupling strength between the resonator and the first cavity.
The cavities are driven by an external source of two-mode squeezed microwave field, which can be produced with a non-degenerate Josephson parametric amplifier \cite{Nat-499-62,Nat-465-64,prl-107-113601,prl-108-123902}. Assuming that the bandwidth of the squeezed microwave field is larger than
the cavity damping rate $\kappa$, then the interaction between the cavity modes  and the external squeezed field is described by \cite{prl-92-013602}
 \begin{eqnarray}
  \mathcal {L}_c\hat{\rho}&=&\kappa M(\hat{a}_1\hat{\rho}\hat{a}_2+\hat{a}_2\hat{\rho}\hat{a}_1-\hat{a}_1\hat{a}_2\hat{\rho}-\hat{\rho}\hat{a}_1\hat{a}_2+\text{H.c.})\nonumber\\
  &&+\frac{\kappa}{2}(N+1)\sum_{j=1,2}(2\hat{a}_j\hat{\rho}\hat{a}_j^\dag-\hat{a}_j^\dag\hat{a}_j\hat{\rho}-\hat{\rho}\hat{a}_j^\dag\hat{a}_j)\nonumber\\
  &&+\frac{\kappa}{2} N\sum_{j=1,2}(2\hat{a}_j^\dag\hat{\rho}\hat{a}_j-\hat{a}_j\hat{a}_j^\dag\hat{\rho}-\hat{\rho}\hat{a}_j\hat{a}_j^\dag).
 \end{eqnarray}
 Here $M$ and $N$ are related to the statistics of the driving broadband two-mode squeezed field: $M$ accounts for the intermode correlations, and $N$ is the average photon number for both modes. For perfect two-mode squeezed vacuum, we have $M=\sinh r_0\cosh r_0$ and $N=\sinh^2 r_0$, with $r_0$ the squeezing parameter.
Therefore, the master equation describing the system dynamics reads
\begin{eqnarray}\label{M1}
 \frac{d\hat{\rho}(t)}{dt}&=&-\frac{i}{\hbar}[\hat{\mathcal{H}},\hat{\rho}]+\mathcal {L}_c\hat{\rho}+\mathcal {L}_m\hat{\rho},
\end{eqnarray}
where
\begin{eqnarray}
\mathscr{L}_m\hat{\rho}&=&\frac{\gamma_\text{m}}{2}(n_\text{th}+1)(2\hat{b}\hat{\rho}\hat{b}^\dag-\hat{b}^\dag\hat{b}\hat{\rho}-\hat{\rho}\hat{b}^\dag\hat{b})\nonumber\\
&&+\frac{\gamma_\text{m}}{2}n_\text{th}(2\hat{b}^\dag\hat{\rho}\hat{b}-\hat{b}\hat{b}^\dag\hat{\rho}-\hat{\rho}\hat{b}\hat{b}^\dag),
\end{eqnarray}
 with $\gamma_\text{m}$ the
mechanical decay rate, and $n_\text{th}$ the thermal phonon number of the mechanical mode.

\noindent\textbf{Enhanced electromechanical coupling via squeezed source driving.}To get more insight into the system's dynamics, it is convenient to introduce two delocalized cavity Bogoliubov modes \cite{pra-88-043802,prl-110-233602,prl-110-253601}
\begin{eqnarray}
\hat{\mathcal {A}}_1&=&\cosh r_0\hat{a}_1+\sinh r_0 \hat{a}_2^\dag\nonumber\\
\hat{\mathcal {A}}_2&=&\cosh r_0\hat{a}_2+\sinh r_0 \hat{a}_1^\dag.
\end{eqnarray}
Using these cavity Bogoliubov modes, the cavity driven term can be rewritten as
 \begin{eqnarray}\label{L1}
  \mathcal {L}_c\hat{\rho}&=&\frac{\kappa}{2}\sum_{j=1,2}(2\hat{\mathcal {A}}_j\hat{\rho}\hat{\mathcal {A}}_j^\dag-\hat{\mathcal {A}}_j^\dag\hat{\mathcal {A}}_j\hat{\rho}-\hat{\rho}\hat{\mathcal {A}}_j^\dag\hat{\mathcal {A}}_j).
 \end{eqnarray}
 This means that for the cavity Bogoliubov modes, the dissipation caused by the system-bath coupling is just like
 that induced by a vacuum bath.
Furthermore, if we choose $r_0=\frac{1}{4}\ln \frac{a+2}{a-2}$, with $a=\frac{\Delta_1+\Delta_2}{\xi}$, then through straightforward derivations,
the system Hamiltonian can be written as
\begin{eqnarray}\label{H1}
  \hat{\mathcal{H}} &=& \hbar \Omega _1 \hat{\mathcal {A}}^\dag_1\hat{\mathcal {A}}_1+\hbar \Omega _2 \hat{\mathcal {A}}^\dag_2\hat{\mathcal {A}}_2+\hbar\omega_\text{m}\hat{b}^\dag\hat{b}\nonumber\\
  &&+\hbar g(\hat{b}^\dag+\hat{b}) (\cosh^2 r_0\hat{\mathcal {A}}_1^\dag\hat{\mathcal {A}}_1+\sinh^2 r_0 \hat{\mathcal {A}}_2 \hat{\mathcal {A}}_2^\dag \nonumber\\
  &&-\sinh r_0\cosh r_0 \hat{\mathcal {A}}_1^\dag\hat{\mathcal {A}}_2^\dag-\sinh r_0\cosh r_0 \hat{\mathcal {A}}_1\hat{\mathcal {A}}_2).
\end{eqnarray}
Here
\begin{eqnarray}
\Omega_1&=&1/2[(\Delta_1+\Delta_2-2\xi)e^{2r_0}+\Delta_1-\Delta_2] \nonumber \\
\Omega_2&=&1/2[(\Delta_1+\Delta_2-2\xi)e^{2r_0}+\Delta_2-\Delta_1].
\end{eqnarray}
In the interaction picture with respect to
\begin{eqnarray}
\hat{\mathcal{H}}_0&=&\hbar \Omega _1 \hat{\mathcal {A}}^\dag_1\hat{\mathcal {A}}_1+\hbar \Omega _2 \hat{\mathcal {A}}^\dag_2\hat{\mathcal {A}}_2+\hbar\omega_\text{m}\hat{b}^\dag\hat{b}
\end{eqnarray}
and under the condition $\Omega_1+\Omega_2\gg g\sinh r_0\cosh r_0,\omega_\text{m}$,
this allows us to selectively activate interaction
terms in the system dynamics as
\begin{eqnarray}\label{H}
  \hat{\mathcal{H}} _\text{eff}&=&\hat{\mathcal{H}}_0+\hbar \mathcal {G}_1(\hat{b}^\dag+\hat{b})\hat{\mathcal {A}}_1^\dag\hat{\mathcal {A}}_1+\hbar \mathcal {G}_2(\hat{b}^\dag+\hat{b})\hat{\mathcal {A}}_2^\dag\hat{\mathcal {A}}_2
\end{eqnarray}
with $\mathcal {G}_1=g\cosh^2 r_0$ and $\mathcal {G}_2=g\sinh^2 r_0$. In the case of $\cosh ^2r_0\gg1$, and $\sinh^2r_0\gg1$, one has $\mathcal {G}_1\simeq \mathcal {G}_2$.
This Hamiltonian describes the electromechanical coupling between the resonator and the cavity Bogoliubov modes $\hat{\mathcal {A}}_j (j=1,2)$ with the effective
coupling strengths $\mathcal {G}_j $. Together with the cavity dissipation described by  (\ref{L1}) and mechanical decay, the system dynamics is just like that
a mechanical resonator simultaneously  couples to two  photonic modes, with photon dissipation caused by a vacuum bath. In this case, the radiation-pressure coupling strength $\mathcal {G}_j$ can be greatly enhanced. By suitably tuning the system parameters $\Delta_1,\Delta_2,\xi$ and the driving source parameter $r_0$, this coupling strength
can be increased by several orders of magnitude to reach the strong-coupling regime, i.e., $\mathcal {G}_j>\kappa,n_\text{th}\gamma_\text{m}$. \emph{This is the main result of this work}: by coupling the mechanical resonator to  cavity Bogoliubov modes, and driving the cavities by a broadband
two-mode squeezed vacuum, the radiation-pressure coupling can be greatly enhanced, even allowed to reach the strong coupling regime.

\begin{figure}[b]
\centerline{\includegraphics[bb=24 196 350 375,totalheight=1.80in,clip]{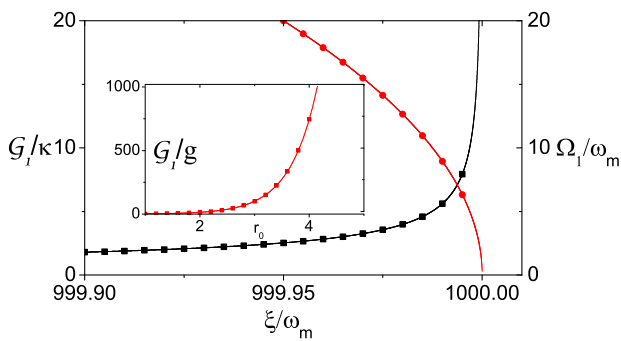}}
\caption{(Color online) The radiation pressure coupling strength $\mathcal{G}_1  $, and the Bogoliubov mode frequency $\Omega_1$ versus the system parameter $\xi$. The relevant parameters are $\Delta_1+\Delta_2=2000\omega_\text{m},g=0.001\omega_\text{m},$ and $\kappa=0.02\omega_\text{m}$.  In the inset  $\mathcal{G}_1$ versus the squeezing parameter $r_0$.  }
\end{figure}
Fig. 2 presents the calculated coupling strength $\mathcal{G}_1$ and the Bogoliubov mode frequency $\Omega_1$ as functions of the system parameter $\xi$.
The results for $\mathcal{G}_2$ and $\Omega_2$ are similar to those for $\mathcal{G}_1$ and $\Omega_1$ as presented in Fig. 2.
As the parameter $\xi$ approaches the critical value $\xi_0=0.5 (\Delta_1+\Delta_2)$, the coupling strength  $\mathcal{G}_1 (\mathcal{G}_2)$ can get a very large value, while the Bogoliubov mode frequency $\Omega_1(\Omega_2)$ tends to be very small. However, there is an optimal point at which both the coupling strength $\mathcal{G}_1 (\mathcal{G}_2)$ and the frequency $\Omega_1(\Omega_2)$ can get a relatively large value, i.e., $\mathcal{G}_1\sim 8\kappa$ and $\Omega_1\sim 8\omega_\text{m}$.  The Bogoliubov mode frequencies $\Omega_1$ and $\Omega_2$ are controllable frequencies, which are determined by the driving frequency detunings $\Delta_1,\Delta_2$ and the parametric pumping strength $\xi$. They should have a relatively large value as compared to the parameters $g\sinh r_0\cosh r_0$ and $\omega_\text{m}$.
In this case we can ignore the parametric amplification terms for the Bogoliubov modes in Eq. (6) and obtain the  standard
electromechanical radiation-pressure interactions displayed  in Eq. (7). This confirms that the single photon strong coupling can be reached with the chosen parameters.

We now consider whether the parameter regime is experimentally accessible with current experimental setups. From
Fig. 2 we find that the scheme works well when $\xi\sim1000 \omega_\text{m}$.
For nanomechanical resonators with frequencies from several kHz to several MHz, the parametrical coupling strength $\xi$ between the cavities is on the order of  hundreds of MHz.
This coupling strength is well accessible with current circuit QED setups \cite{prl-114-090503}.  The value of $g\sim 0.001 \omega_\text{m}$
is an order of magnitude lager than that achieved via mechanical resonators with frequencies of several MHz  in current technology \cite{Nat-471-204}.
However, by the usage of mechanical  resonators with low frequencies of hundreds of kHz, the value of $g\sim 0.001 \omega_\text{m}$ is accessible in
present experiments \cite{NC-3-987}.
From the inset of this figure, we also note that this proposal
requires a two-mode squeezing source with a very large squeezing parameter $r_0$.  Though this requirement
is somewhat challenging, it is within reach of the state-of-the-art experiments. Recent experiments in two-mode squeezing of microwave radiation rely on the amplification of quantum noise by the Josephson Parametric Converter \cite{Nat-499-62,Nat-465-64,prl-107-113601,prl-108-123902}. It has been demonstrated that the generated two-mode squeezing source can possess the highest gain of $\cosh^2(r_0)=-51$ dB, which corresponds to a squeezing parameter $r_0\sim 6.5$. Therefore, this protocol can be realized with the state-of-the-art techniques of superconducting integrated circuits and electromechanical systems.

\noindent\textbf{Applications.}The effective strong coupling offers great potential
for single-photon manipulation and quantum states generation.
For example, the photon blockade phenomenon  existing  in the strongly coupled optomechanical system can occur
in this system. This is quantitatively characterized by
the zero-delay second-order correlation function
\begin{eqnarray}
g^{(2)}_i(0)&=&\langle \hat{\mathcal{A}}_i^\dag\hat{\mathcal{A}}_i^\dag\hat{\mathcal{A}}_i\hat{\mathcal{A}}_i\rangle(t)/\langle \hat{\mathcal{A}}_i^\dag\hat{\mathcal{A}}_i\rangle^2(t).
\end{eqnarray}
This quantity provides a direct experimental measure for nonclassical
antibunching effects if $g^{(0)}_i<1$   and for $g^{(0)}_i\rightarrow0$   indicates a full photon blockade for the $i$th Bogoliubov mode.

We now discuss the feasibility of the measurement of the photon statistics in experiments. A challenge here is that the Bogoliubov modes are a linear combination of the two cavity modes, and are more difficult to be addressed separately. We assume the first cavity is weakly driven by a probe field with frequency $\omega_\text{p}$ and amplitude $\mathcal {E}$. The Hamiltonian
is given by
\begin{eqnarray}
\hat{\mathcal {H}}_\text{p}&=&\mathcal{E}(\hat{a}_1e^{i\omega_\text{p}}+\hat{a}_1^\dag e^{i\omega_\text{p}}).
\end{eqnarray}
In the frame rotating at half the
flux driving frequency $\omega_\text{d}$ and in the Bogoliubov mode representation, we have
\begin{eqnarray}
 \hat{\mathcal {H}}_\text{p}=\mathcal{E}[\cosh r_0\hat{\mathcal{A}}_1e^{i\omega'_\text{p}}-\sinh r_0 \hat{\mathcal{A}}_2^\dag e^{i\omega'_\text{p}}+\text{H.c.}]
\end{eqnarray}
with $\omega'_\text{p}=\omega_\text{p}-\omega_\text{d}/2$. Under the condition $\omega'_\text{p}\simeq \Omega_1,\Omega_2$, $\omega'_\text{p}+\Omega_2\gg \mathcal{E}\sinh r_0,\mathcal{E}\cosh r_0,\omega_\text{m}$, we can safely ignore the rapidly oscillating terms under the rotating-wave approximation
\begin{eqnarray}
\hat{\mathcal {H}}_\text{p}=\mathcal{E}[\cosh r_0\hat{\mathcal{A}}_1e^{i\omega'_\text{p}}+\text{H.c.}].
\end{eqnarray}
Therefore, with suitably choosing the
probe parameters, one can selectively address the Bogoliubov modes on demand. The measurement of the photon statistics needs monitoring the signal coming out from the system and  it is not known how the detailed measurement goes.

\begin{figure}[t]
\centerline{\includegraphics[bb=176 14 905 631,totalheight=2.8in,clip]{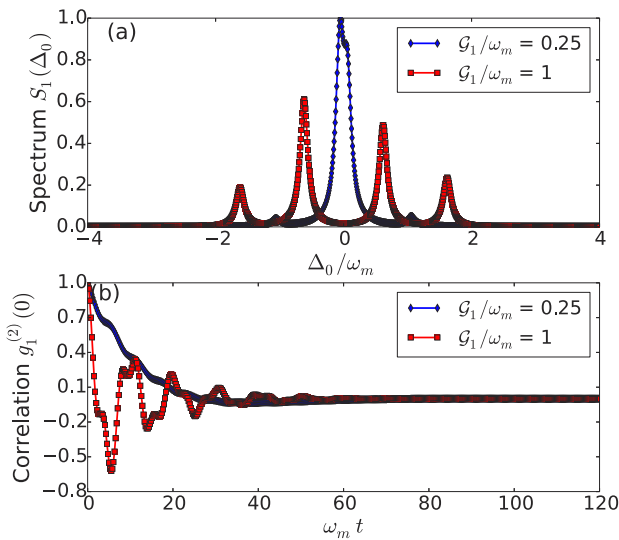}}
\caption{(Color online) (a) Cavity excitation spectrum $S_1(\Delta_0)$ with
different values of the coupling strength $\mathcal {G}_1/\omega_\text{m}$ in the resolved sideband regime $\kappa\sim 0.1\omega_\text{m}$.
(b)  The second-order correlation function $g^{(2)}_1(0)$ versus time. In both
plots, $T=0$ and $Q=10^3$. }
\end{figure}
In Fig.3 (a), we plot  the excitation spectrum for the  first Bogoliubov mode
\begin{eqnarray}
 S_1(\Delta_0)&=&\lim_{t\rightarrow\infty}\langle  \hat{\mathcal{A}}_1^\dag\hat{\mathcal{A}}_1\rangle/n_0,
\end{eqnarray}
 obtained by numerically solving Eq (\ref{M1}),  as a function of the detuning $\Delta_0=\Omega_1-\omega'_\text{p}$ for different values of the coupling
strength $\mathcal {G}_1/\omega_\text{m}$, with $n_0=\mathcal{E}^2\cosh^2r_0/\kappa^2$. We observe in the
resolved sideband regime  $\kappa\ll\omega_\text{m}$, a redshift of
the zero phonon line (ZPL) towards $\Delta_0=-\mathcal {G}_1^2/\omega_\text{m}$  and the
appearance of additional resonances at multiples of the
mechanical frequency. These peaks result from
phonon-assisted excitation processes,  providing
a clear indication for single-photon strong coupling optomechanics.
In Fig.3 (b) we present numerical results for the second-order correlation function $g^{(2)}_1(0)$ versus time
through solving the master equation (\ref{M1}) with the transformed Hamiltonian (\ref{H1}). It is clear that the steady state value
for $g^{(2)}_1(0)$ approaches zero for the first Bogoliubov mode, which is a strong
signature of photon blockade.

\noindent\textbf{Discussion and conclusion.}

\noindent In this work, we focus on implementing the idea using two parametrically coupled superconducting cavities in
an electromechanical system. The usage of two cavities rather than one is particularly appealing when one considers applying
this proposal to quantum information processing such as distributed quantum computation and quantum network. This proposal can also apply to a single microwave cavity capacitively coupled to a mechanical resonator. In this case, it will need a  single-mode squeezed field to drive the cavity, and a superconducting qubit in the cavity to produce the desired nonlinearity for the cavity mode \cite{pra-86-013814}.

To conclude, we have proposed an efficient method for enhancing the radiation-pressure type
coupling in an electromechanical system. We have shown that, by driving the cavities with a two-mode squeezed microwave source and selectively coupling the resonator to the cavity Bogoliubov modes, the effective interaction strengths between the mechanical resonator and the cavities can  be enhanced into
the single-photon strong-coupling regime. This allows the investigation of a number of interesting phenomena such as photon blockade effects and the generation of
quantum states  with  currently available technology in the field of cavity  electromechanics.

\noindent\textbf{Acknowledgements}

\noindent This work was supported by the National Nature Science Foundation of China (
Grants No. 11474227, No. 11174233, and No. 11374239), and the Fundamental Research
Funds for the Central Universities.
Part of the simulations are coded in PYTHON using the QUTIP library.

\noindent\textbf{ Author Contributions}

\noindent  P. B. L. conceived the idea and carried out the calculations. H. R. L., and F. L. L. participated in the discussions. All authors contributed to the interpretation of the work.

\noindent\textbf{  Additional information}

\noindent Competing Financial interests: The authors declare no competing financial interest.

\end{document}